\documentclass[useAMS,usenatbib,letterpaper]{mn2e}
\usepackage[varg]{txfonts}
\usepackage{graphicx}
\usepackage{natbib}
\usepackage{hyperref}
\hypersetup{
  colorlinks   = true,    
  urlcolor     = blue,    
  linkcolor    = blue,    
  citecolor    = blue      
}
\usepackage[usenames,dvipsnames]{xcolor}
\usepackage[T1]{fontenc}
\usepackage{aecompl} 

\def\apjl{ApJL }

\def\aj{AJ }

\def\apj{ApJ }
\def\pasp{PASP }

\def\apjs{ApJS }

\def\mnras{MNRAS }
\def\aap{A\&A }

\begin{document}

\title[NACO/APP A and F Main Sequence Star Survey]
{Searching for gas giant planets on Solar System scales - A NACO/APP $L'$-band 
survey of A- and F-type Main Sequence stars}

\author[T. Meshkat et al.]
  {T.~Meshkat,$^1$\thanks{Based on observations collected at the European 
  Organization for Astronomical Research in the Southern Hemisphere, 
Chile, ESO under program numbers 089.C-0617(A), 089.C-0149(A).}
M.~A.~Kenworthy,$^1$ M.~Reggiani,$^2$ S.~P.~Quanz,$^2$
E.~E.~Mamajek,$^3$ \newauthor M.~R.~Meyer$^2$ \\
  $^1$Leiden Observatory, P.O. Box 9513, Niels Bohrweg 2,
  2300 RA Leiden, The Netherlands\\
  $^2$Institute for Astronomy, ETH Zurich,
  Wolfgang-Pauli-Strasse 27, 8093 Zurich, Switzerland\\
  $^3$Department of Physics and Astronomy, University of
  Rochester, Rochester, NY 14627-0171, USA}
\date{2015 March 17}

\pagerange{\pageref{firstpage}--\pageref{lastpage}} \pubyear{2015}

\maketitle

\label{firstpage}

\begin{abstract}
We report the results of a direct imaging survey of A- and F-type main sequence stars 
searching for giant planets. A/F stars are often the targets of surveys, as they are 
thought to have more massive giant planets relative to solar-type stars. However, 
most imaging is only sensitive to orbital separations $>$\,30 AU, where it has been 
demonstrated that giant planets are rare. In this survey, we take advantage of the 
high-contrast capabilities of the Apodizing Phase Plate coronagraph on NACO at the 
Very Large Telescope. Combined with optimized principal component analysis 
post-processing, we are sensitive to planetary-mass companions (2 to 12 
$M_{\rm Jup}$) at Solar System scales ($\leq$30 AU). We obtained data on 13 stars 
in $L'$-band and detected one new companion as part of this survey: an M$6.0\pm0.5$ 
dwarf companion around HD 984. We re-detect low-mass companions around HD 
12894 and HD 20385, both reported shortly after the completion of this survey. We use 
Monte Carlo simulations to determine new constraints on the low-mass ($<$80 
$M_{\rm Jup}$) companion frequency, as a function of mass and separation. Assuming 
solar-type planet mass and separation distributions, normalized to the planet frequency 
appropriate for A-stars, and the observed companion mass-ratio distribution for stellar 
companions extrapolated to planetary masses, we derive a truncation radius for the 
planetary mass companion surface density of $<$135 AU at 95\% confidence. 
\end{abstract}


\section{Introduction}
Stellar properties are an important metric in the search for planets, as they guide the 
target selection for detection surveys. In particular, stellar mass and metallicity are 
significant quantities in determining both the formation and evolution of stars and 
planets \citep{Johnson10}. Several radial velocity (RV) studies have shown that the 
giant planet frequency increases with stellar metallicity \citep{Santos04,Fischer05}. 
The giant planet population as a function of stellar mass, however, is not consistent 
between different planet detection techniques \citep{Quanz12,Clanton14a,Vigan12}. 
While progress has been made in linking the RV and microlensing populations 
\citep{Clanton14a}, this is a challenging problem involving the synthesis of different 
biases and parameter spaces covered by all the detection techniques.

Gas giant planets ($>1\,M_{\rm Jup}$) are the only directly imaged planets thus far, 
due to their increased self-luminous thermal emission and decreased contrast at 
infrared wavelengths with the star. Planet populations derived from RV surveys are 
often extrapolated to larger orbital separations to analyze the frequency of giant 
planets in direct imaging surveys (e.g. \citealt{Lafreniere07,Biller13}).

Planet formation scenarios \citep{Alibert11} and simulations extrapolating RV planet 
populations \citep{Crepp11, Johnson07} suggest that massive stars 
(${>}1.3 M_{\sun}$) are the most favorable targets for directly imaging planets, since 
they have proportionally more material to form giant planets. Indeed many directly 
imaged planetary mass companions have been found around A or F stars: HR8799 
bcde \citep{Marois08,Marois10}, $\beta$ Pic b \citep{Lagrange09, Lagrange10}, HD 
95086 b \citep{Rameau13b,Rameau13c}, HD 106906 b \citep{Bailey14}. The 
detection of the HR8799 planets was the result of the \citet{Vigan12} International 
Deep Planet Survey. Most surveys, however, have yielded null results 
\citep{Desidera15, Chauvin15, Janson13, Rameau13, Biller13, Chauvin10, Heinze10, 
Lafreniere07, Kasper07}. These null results are likely due to the lack of contrast at 
small orbital separations, where most planets are expected to be found. Typical 
detection limits for these surveys are 5-20 $M_{\rm Jup}$ for $>30$ AU. Planets are 
rare at large orbital separations \citep{Chauvin10, Lafreniere07, Nielsen10} but at 
Solar System scales ($\leq30$ AU), stars are largely unexplored. 

The main limitations for direct imaging are stellar ``speckles'' which can appear 
brighter than a companion \citep{Hinkley09}. Coronagraphs are used in order to reach 
smaller angular separations around stars. They reduce the diffraction due to scattered 
stellar light in the telescope optics but at a cost of reduced throughput. The Apodizing 
Phase Plate (APP; \citealt{Kenworthy10b,Quanz10,Quanz13}) coronagraph 
suppresses the diffraction in a 180$^{\circ}$ wedge around a star, increasing the 
chances of detecting a very close-in companion. Several studies have demonstrated 
the APP's capability of reaching $\leq30$ AU 
(\citealt{Meshkat15,Kenworthy13,Quanz11}). 

We aim to probe down to Solar System scales ($\leq30$ AU) around 13 A- and F-type 
main sequence stars in order to detect giant planets as well as set constraints on the 
planet frequency. We use the APP coronagraph on NACO at the Very Large Telescope 
(VLT), the $L'$-band filter, and optimized Principal Component Analysis (PCA) to 
achieve deep sensitivity limits (2 to 10 $M_{\rm Jup}$ at $\leq30$ AU). 

In Section \ref{sec:obs} we describe our target selection process, the coronagraphic 
observations, our data reduction method and how we determine the sensitivity of our 
data. In Section \ref{sec:discussion} we discuss the sensitivity achieved, our new 
detection of an M$6.0\pm0.5$ dwarf companion to HD 984, and our re-detection of 
companions to HD 12894 and HD 20385. We run Monte Carlo simulations to 
determine the probability distribution of our results, in order to compare different planet 
population models for A-type stars. Our conclusions are in Section \ref{sec:conclusion}.

\section{Observations and Data Reduction}
\label{sec:obs}

Our sample was carefully selected to derive the best possible constraints on the 
frequency of giant exoplanets on Solar System scales: nearby, young, and massive 
main sequence stars. Young planets are still warm from their contraction 
\citep{Spiegel12}. By converting gravitational energy into luminosity, they are bright in 
the infrared. However, determining the age of a main-sequence star can be extremely 
challenging. One way to deal with this difficulty is to only select targets which are 
members of nearby associations with well established ages. If they are all bona-fide 
members of the group, we can assume the stars are of a similar age. Except for 
one\footnote{At the time of our observations, HD 984, was believed to be a 30 Myr 
member of Columba association. Based on our detection of a low-mass stellar 
companion to HD 984 and independent isochrone fitting, we estimate the age of HD 
984 to be 115\,$\pm$\,85 Myr (Meshkat et al. \textit{accepted}).}, our targets are all 
members of nearby young moving groups or associations: $\beta$ Pic Moving Group 
\citep[23\,$\pm$\,3 Myr;][]{Mamajek14}, Tuc-Hor Association \citep[40 Myr;][]{Kraus14},
AB Dor Association \citep[125\,$\pm$\,15 Myr;][]{Barenfeld13}. Nearby stars allow us 
to search for companions at smaller physical separations. We aim to reach planet 
sensitivity on Solar System scales, where we expect more planets to reside 
\citep{Chauvin10,Lafreniere07,Nielsen10}. Thus, we have selected only stars which 
are less than 66 pc away.

At the time of selection, most\footnote{HD 20385 was known to have a wide binary 
12$\arcsec$ away.} of the targets were known to be single stars 
\citep{Mason11,Pourbaix09} and not in the denser nucleus of their association. 
However, shortly after our survey was completed, a companion was discovered 
around one of our targets, HD 12894, by \citet{Biller13} and \citet{Rameau13}. 
Another target, HD 20385, was found to have a companion shortly after our data 
were acquired \citep{Hartkopft12}.

\subsection{Observations at the VLT}

Data were obtained for 13 targets from 2011 to 2013 (088.C-0806(B), 089.C-0617(A) 
PI: Sascha Quanz) at the Very Large Telescope (VLT)/UT4 with NACO 
(\citealt{Lenzen03,Rousset03}) and the APP coronagraph \citep{Kenworthy10b}. The 
variation in the observing time for each target depends on the observing conditions 
on the night; if the observing conditions fell below a threshold during the night, the 
data acquisition was cancelled. Data were obtained with the L27 camera, in the 
$L'$-band filter ($\lambda$ = 3.80$\mu$m and $\Delta\lambda$\,=\,0.62$\mu$m) and 
the NB 4.05 filter ($\lambda$ = 4.051$\mu$m and $\Delta\lambda$\,=\,0.02$\mu$m) 
depending on the star's $L'$-band magnitude. The visible wavefront sensor was used 
with each target star as its own natural guide star. We observed in pupil tracking 
mode to perform Angular Differential Imaging (ADI; \citealt{Marois06}). We 
intentionally saturated the point spread function (PSF) core (on average out to 
$\sim$0\farcs08) to increase the signal-to-noise (S/N) from potential companions in 
each exposure. Unsaturated data were also obtained to calibrate photometry relative 
to the central star.

The APP generates a dark D-shaped wedge on one half of a target. Excess scattered 
light is increased on the other side of the target, which is not used in the data analysis. 
Two datasets were obtained with different initial position angles (P.A.) for full 
360$^{\circ}$ coverage around the target star. Data were obtained in cube mode.
\autoref{table:stellar_properties} lists the stellar properties for each of our targets.
\autoref{table:data} lists the observing conditions for all the data obtained. 13 targets 
were observed in at least one APP hemisphere.

\begin{table}
\scriptsize
\caption{Overview of stellar values used for each target.}
\centering 
 \begin{tabular}{lccccc}
\hline
   Target & Mass ($M_{\odot}$) & $L'$ mag  & Spectral type  & Distance (pc) & Age (Myr)\\ 
  \hline
HD 203 & 1.40 & 5.2 & F3V & 39.4\,$\pm$\,0.6 & 23 \\ 
   
HD 12894 & 1.39 & 5.5 & F4V & 47.8\,$\pm$\,1.0 & 40 \\ 
   
HD 25457 & 1.21 & 4.3 & F6V & 18.8\,$\pm$\,0.1 & 125 \\ 
   
HD 35114 & 1.16 & 6.2 & F6V & 8.3\,$\pm$\,0.9 & 40 \\ 
   
HD 20385 & 1.13 & 6.4 & F6V & 49.2\,$\pm$\,1.5 & 40 \\ 
 
HD 102647 & 1.9 & 1.9 & A3Va & 11.0\,$\pm$\,0.1 & 40 \\ 
 
HD 984 & 1.18 & 6.0 & F7V & 47.1\,$\pm$\,1.4 & 115\,$\pm$\,85\\

HD 13246 & 1.18 & 6.2 & F7V & 44.2\,$\pm$\,0.9 & 40 \\ 
 
HD 40216 & 1.24 & 6.2 & F7V & 54.4\,$\pm$\,1.3 & 40 \\ 
   
HD 30051 & 1.38 & 6.0 & F2/3IV/V & 63.6\,$\pm$\,4.2 & 40 \\ 
   
HD 25953 & 1.16 & 6.6 & F5 &  55.2\,$\pm$\,2.9 & 125 \\ 

HD 96819 & 2.09 & 5.2 & A1V & 55.6\,$\pm$\,1.7 & 23 \\ 

HD 123058 & 1.30 & 6.7 & F4V & 64.1\,$\pm$\,3.5 & 40 \\ 
   \hline
   \end{tabular}  

   \medskip
Distances are extracted from parallaxes in the Hipparcos catalog 
\citep{vanLeeuwen07}. The $L'$-band mag is converted from $K$-band mag in the 
2MASS survey \citep{Cutri03} to using \citet{Cox00}. The masses are from 
\citet{Casagrande11} and \citet{Chen14}. All ages are based on membership in 
nearby young moving groups or associations, taken from 
\citet{Mamajek14,Kraus14,Luhman05,Barenfeld13}, except for HD 984 which we 
compute in Meshkat et al. \textit{accepted}. The bottom six targets were only observed 
in one APP hemisphere. 
\label{table:stellar_properties}
\end{table}

\begin{table*}
\small
\caption{Observing Log for NACO/VLT 088.C-0806(B) and 089.C-0617(A) } 
\centering 
\begin{tabular}{l p{0.17\linewidth} cccc}
\hline 
   Target & Observation dates UT  (Hem 1, Hem 2) & Number of data cubes & Total integration time (min) 
   & On-sky rotation ($^{\circ}$)  & Average DIMM seeing ($\arcsec$)\\
  \hline
   HD 203 & 2011 Oct 12, 2011 Nov 07 & 154, 64 & 76.8, 58.0 & 46.65, 60.43 &  0.90, 1.19 \\ 
   
   HD 12894 & 2011 Dec 10, 2011 Dec 24 & 65, 55 & 55.6, 47.1 & 31.97, 27.92  & 0.78, 1.23 \\  
   
   HD 25457 & 2011 Dec 11, 2011 Dec 21 & 71, 65 & 43.4, 39.8 & 28.63, 25.92 & 1.44, 0.65 \\
   
   HD 35114 & 2011 Dec 13, 2012 Jan 02 & 35, 35 & 30.3, 30.3 & 31.98, 28.20  & 0.75, 1.02 \\
   
   HD 20385 & 2011 Dec 21, 2012 Jan 08 & 47, 35 & 40.4, 30.3 & 30.54, 24.61 & 1.82, 1.25 \\

   HD 102647 & 2012 Jun 01, 2013 Apr 26 & 47, 56 & 47.3, 56.2 & 20.34, 24.71 &  1.61, 0.76 \\
 
   HD 984 & 2012 Jul 18, 2012 Jul 20 & 59, 65 & 54.5, 59.9 & 47.41, 42.46 &  0.65, 0.86 \\

   HD 13246 & 2011 Dec 07, -- & 107, -- & 90.9, -- & 45.43, -- &  0.99, -- \\     
 
   HD 40216 & 2012 Jan 03, -- & 11, -- & 10.1, -- & 11.74, -- &  0.88, -- \\
   
   HD 30051 & 2012 Jan 07, -- & 56, -- & 48.0, -- & 2.7, -- &  1.68, --  \\
   
   HD 25953 & 2012 Jan 13, -- & 79, -- & 67.3, -- & 42.76, -- &  0.77, -- \\ 

   HD 96819 & 2012 May 02, -- & 49, -- & 47.3, -- & 105.46, -- &  0.64, -- \\

   HD 123058 & 2012 May 21, -- & 62, -- & 63.5, -- & 29.44, -- &  0.77, -- \\
   \hline
   \end{tabular}  
   
    \medskip
Data are in chronological order based on first Hemisphere observed. The last six
targets were only observed in one APP hemisphere. Targets are listed in the same 
order as Table \ref{table:stellar_properties}. 
\label{table:data}
\end{table*}

\subsection{Data Reduction}
\label{sec:data_reduction}

A dither pattern on the detector was used to subtract sky background and detector 
systematics from the raw data, as detailed in \citet{Kenworthy13}. Subtracted data 
cubes are centroided and averaged over. The two APP hemispheres obtained for 
each target must be processed separately, since they were observed on different 
nights and thus have different speckle noise patterns. Optimized PCA was run on 
both of the APP hemispheres independently for each target, following 
\citet{Meshkat14}. This involves creating a linear combination of principal components 
(PCs) from the data itself in order to model and subtract away the stellar diffraction. 
Only the 180$^{\circ}$ D-shaped dark hemisphere was used in the PCA analysis. We 
fixed the number of PCs at approximately 10\% the number of input frames, as this 
yields the optimal PSF subtraction close ($<1\farcs0$) to the star. We searched for 
point sources using this method for all 13 of our targets, despite not having full 
360$^{\circ}$ coverage for 6 of them.

We injected fake planets into our data (before PCA processing) in order to determine 
the 5$\sigma$ sensitivity limit for each target. Unsaturated data of the star was used to 
inject the fake planets. We scaled the unsaturated data to the same exposure as the 
saturated data. The star was added to the data with a contrast of 5 to 12 mag in steps 
of 1 mag and from $0\farcs18$ to $1\farcs36$ in steps of $0.13$. The outer radius limit 
was chosen because the field-of-view (FOV) of the APP is limited to only the upper 
quarter of the detector \citep{Kenworthy10b}. The planet injected data was processed 
with PCA, de-rotated, and averaged over for the final image with North facing up.

\begin{figure}
 \centering
\includegraphics[width=84mm]{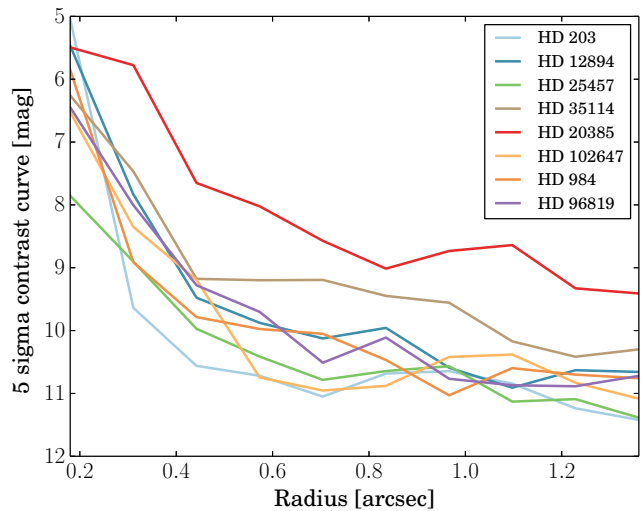}
\caption{5$\sigma$ contrast curves for the targets with full $360^{\circ}$ APP coverage 
around the star and one target, HD 96819, with nearly 360$^{\circ}$ coverage. }
\label{fig:cc}
\end{figure}

The final image was smoothed by a $\lambda$/D aperture, in order to remove features 
which are not the expected planet size or shape \citep{Amara12, Bailey13}. We define 
the S/N of the injected planet to be the value of a single pixel at the location of the 
planet divided by the root mean square (rms) of a ring around the star at the angular 
separation of the planet, excluding the planet itself. Only the statistically independent 
pixels (one smoothing kernel apart) were used to compute the rms. \autoref{fig:cc} 
shows the 5$\sigma$ contrast curves for all 7 targets with two APP hemispheres and one 
target (HD 96819) with nearly full sky coverage in one APP 
hemisphere\footnote{HD 96819 has on sky rotation of $105\fdg46$. Since we are in ADI 
mode, the $180^{\circ}$ APP ``dark hole'' region rotates on the sky and only a 
$\sim74\fdg9$ wedge is missing.}. For targets with two APP hemispheres, fake planets 
were added at a fixed P.A. in each hemisphere. The average S/N of the injected fake 
planets in each hemisphere is used (at the same separation). In the overlapping region 
between the two hemispheres, the number of frames varies slightly. However, since these 
regions are small and the number of frames never varies by more than 20\%, the impact on 
the contrast curves is small.

On average, we achieved a contrast of 9 mag at $0\farcs4$ and 10 to 11 mag at $>0\farcs6$. 
The decreased sensitivity at $<0\farcs7$ around HD 20385 is an outlier compared to the 
other targets. One possible explanation is the extremely bright companion detected at 
$\sim0\farcs8$ (discussed in Section \ref{sec:HD20385}), which affects the PCA component 
determination.

\begin{figure}
 \centering
\includegraphics[width=84mm]{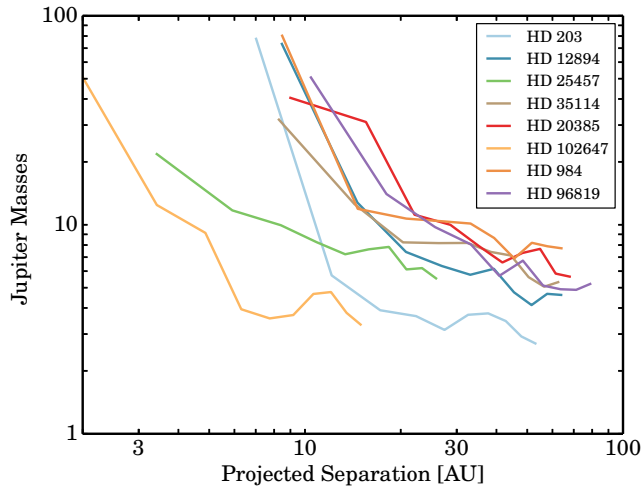}
\caption{Detection limits in Jupiter masses versus projected separation in AU for all targets 
with full APP coverage (and HD 96819). We are sensitive to planet mass companions 
($<12 M_{\rm Jup}$) for all of our targets. HD 102647 is more sensitive at 
smaller projected separation because it is much closer than the rest of the targets (11 pc). 
We do not plot the sensitivity beyond $1\farcs5$ for each target, due to the limited FOV of 
the APP. }
\label{fig:cc_AU}
\end{figure}

We used the COND evolutionary tracks \citep{Baraffe03} to convert the contrast curves to 
planet mass detection limits (\autoref{fig:cc_AU}). We were sensitive to 
planetary mass objects ($< 12\,M_{\rm Jup}$) at different projected separations, depending 
on the target distance. The outer radius for the sensitivity curves were based on the limited 
FOV of the APP \citep{Kenworthy10b}. Thus, while the sensitivity curves appear to flatten 
out, we cannot extend these curves beyond $1\farcs5$ since we were not sensitive 
completely around the star.

\section{Results}
\label{sec:discussion}
We detect one new M$6.0\pm0.5$ dwarf companion to HD 984, and re-detect companions 
to HD 12894 and HD 20385. To estimate the astrometry, we first determined the centroid of 
the three companions. In this way, we verified that our star was well centered in our data. 
Since the APP has an asymmetric PSF, this step is crucial. We then injected fake negative 
companions at the location of the companion to determine the photometry and astrometry 
with error bars.  We also varied the flux of the fake negative companions to cancel out the 
companion flux, which in some cases varies up to 20\% due to atmospheric fluctuations. 

We iteratively converged on the P.A., angular separation, and $\Delta$ magnitude by 
varying the position and contrast and taking a $\chi^{2}$ minimization over the $\lambda/D$ 
aperture at the location of the companion. For the very bright companions to HD 12894 and 
HD 20385, we determined the photometry and astrometry from ADI alone rather than PCA. 
Minor variations in the brightness and position of the companion in each frame can lead to 
``striping'' in the final PCA processed image. This striping occurs when PCA fits the 
remaining flux around of the companion after the fake companion is subtracted, since we 
never perfectly subtract the companion in individual frames due to seeing variability. Table 
\ref{table:companions} lists the properties of the companions we detect based on our APP 
data. The error on the P.A. includes uncertainties from true North orientation, based on 
direct imaging observations ($\sim 0.5^{\circ}$, \citealt{Rameau13}).

\begin{table*}
\caption{Companion properties of our targets.} 
\centering 
\begin{tabular}{l c c c c c }
\hline 
   Target & Date & Separation (arcsec)  & P.A. (deg)  & $\Delta L'$ (mag)) & Mass ($M_{\odot}$)\\
  \hline
   HD12894 B& 2011 Nov 24 & 0.31\,$\pm$\,0.01 & 240.24\,$\pm$\,1.27 & 2.89\,$\pm$\,0.14 & 0.53\,$\pm$\,0.04\\ 
   
   HD20385 B& 2011 Dec 21  & 0.87\,$\pm$\,0.01 & 118.67\,$\pm$\,0.49 & 2.52\,$\pm$\,0.10 & 0.33\,$\pm$\,0.03\\ 

   HD984 B& 2012 July 18  & 0.19\,$\pm$\,0.02 & 108.9\,$\pm$\,3.1 & 6.0\,$\pm$\,0.2 & 0.11\,$\pm$\,0.01\\ 
\hline
   
   \end{tabular}  
\label{table:companions}
\end{table*}

\begin{figure}
\centering
\includegraphics[width=84mm]{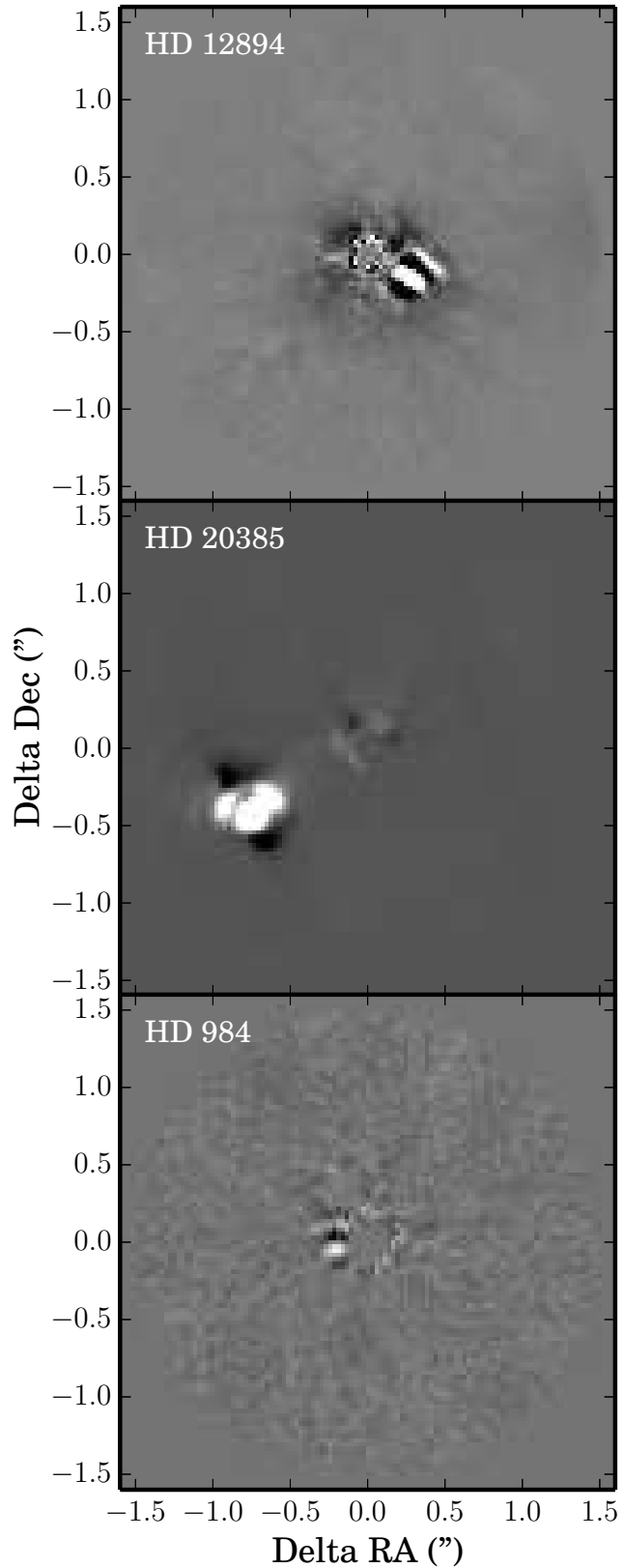}
\caption{Top: HD 12894 ADI processed image, with the companion at $0\farcs31$. 
Middle: HD 20385 ADI processed image, with the companion at $0\farcs87$. The shape of 
the companion PSF is due to the APP PSF.
Bottom: HD 984 PCA processed image with the companion $0\farcs19$ from star.}
\label{fig:companions}
\end{figure}

\subsection{HD 12894}
\citet{Rameau13} and \citet{Biller13} reported the discovery of HD 12894 B. 
\citet{Rameau13} detected a $\Delta L'=2.7\,\pm\,0.1$ mag point source 14 AU from the star.  
They concluded that it was likely bound to the star, based on a non-detection in 1999 2MASS 
data, because a background source would be detected due to proper motion of HD 12894. 
Based on the contrast and an age of 30 Myr, they concluded that the companion was a 
$0.8 M_{\odot}$ K6 star. \citet{Biller13} concluded it was a co-moving 0.46\,$\pm$\,0.08 
$M_{\odot}$ companion 15.7\,$\pm$\,1.0 AU away from its star, based on contrast of 
$\Delta H=3.0$ mag. The discovery was made with the VLT/NACO instrument in $L'$-band 
as well as the NICI instrument on Gemini South in $H$-band, respectively.

At the time our data were acquired, the companion was not known. We detect the companion 
with a $\Delta L'=2.89\,\pm\,0.14$ mag at a PA of 240.24\,$\pm\,1\fdg27$ and separation of 
$0\farcs31\,\pm\,0\farcs01$. This corresponds to a projected separation of 14.8\,$\pm$\,0.8 
AU, with d=47.8\,$\pm$\,1.0 pc for the distance to the star \citep{vanLeeuwen07}. 
\autoref{fig:companions} shows the ADI processed image of the companion. Using the COND 
evolutionary models \citep{Baraffe03} and the age of the star (40 Myr: \citealt{Kraus14}), we 
estimate the mass of the companion is 0.53\,$\pm$\,0.04 $M_{\odot}$. Our analysis of this 
companion is consistent with \citet{Biller13}.

\subsection{HD 20385}
\label{sec:HD20385}
At the time we proposed to observe this target, it was a known wide binary, with a companion 
TOK 78 B 12$\arcsec$ away. \citet{Hartkopft12} reported the discovery of a new, close 
companion around HD 20385 at $0\farcs88$, $\Delta I = $3.5 mag, $\Delta y =$ 5.2 mag. 
They estimated the companion has a period of 200 years. Due to its variable radial velocity, 
they suggested the companion could be two unresolved companions.

We detect the companion in our data at $0\farcs87\,\pm\,0\farcs01$ with a P.A. of 
$118\fdg67\,\pm\,0\fdg49$ and contrast of $\Delta L'=2.52\,\pm\,0.10$ mag. Using the stellar 
distance of 49.2\,$\pm$\,1.5 pc \citep{vanLeeuwen07}, this companion is at a projected 
separation of 42.8\,$\pm$\,1.8 AU. \autoref{fig:companions} shows the ADI processed image 
of the companion to HD 20385. The companion's PSF clearly shows the APP PSF structure, 
with the bright lobes smeared due to the rotation on the sky. The age of the system is 40 Myr 
based on membership in the Tuc-Hor Association \citep{Kraus14}. Using COND models 
\citep{Baraffe03}, we estimate the companion to have a mass of 0.33\,$\pm$\,0.03 $M_{\odot}$.

\subsection{HD 984}

As reported in Meshkat et al. (\textit{accepted}), we detected a companion around HD 984 in 
our APP data, as well as archival direct imaging data. \autoref{fig:companions} shows our PCA 
reduced image with 20 PCs. We confirmed the companion is co-moving with HD 984 and 
determined that is an M$6.0\pm0.5$ dwarf based on SINFONI integral field spectroscopy (see 
Table \ref{table:companions}).

\subsection{Monte Carlo Simulations}
\label{sec:monte_carlo}
\begin{figure}
 \centering
\includegraphics[width=84mm]{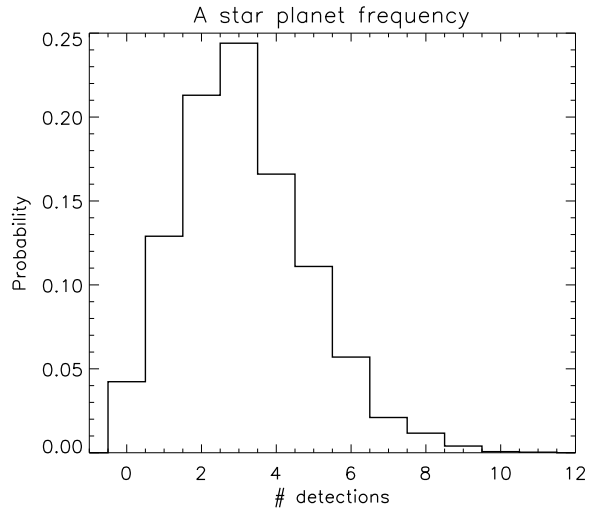}\\
\caption{Detection probability distributions for our survey, assuming an 
outer-radius cutoff of 140 AU for the planet separation distributions 
(Reggiani et al., \textit{submitted}). 
With our null result, we rule out the A-star frequency beyond a cutoff of 135 AU 
at the 95\% confidence level. 
}
\label{fig:prob_dist}
\end{figure}

We ran 10000 Monte Carlo simulations of the target stars in order to determine the probability 
distribution of detecting substellar companions in our data assuming power-law slopes for the 
mass and semi-major axis distributions, for both planets and brown dwarfs (BDs), following 
Reggiani et al. \textit{submitted}. For the BD distribution, we adopted the stellar companion 
mass ratio distribution (CMRD) from \citet{Reggiani13} and a log-normal separation 
distribution \citep{DeRosa14}. We included all of our targets in these simulations, including 
those with only one APP hemisphere coverage. In order to account for the targets without full 
sky coverage in our simulations, we multiplied the overall planet frequency per target by the 
fraction of sky coverage achieved.

For the planets, we adopted a model distribution for A-type primaries which assumes the 
mass and separation distributions measured by RV surveys for solar-type stars, and 
extrapolated at larger separations (\citealt{Heinze10}, based on \citealt{Cumming08}). 
According to \citet{Cumming08}, the planet mass and semi-major axis distributions are 
modeled with power laws of index $\alpha$=-1.31 and $\beta$=-0.61 ($dN\sim M^{-1.31}dM$ 
and $dN\sim a^{-0.61}da$). To account for the A-type primary planet frequency, we adopted 
the values corresponding to the median sensitivity achieved in the RV data presented in 
\citet{Bowler10} and \citet{Johnson10} (f=$11\pm2\%$ for planets in the ranges 0.5-14 
$M_{\rm Jup}$ and 0.1-3.0 AU).

For both planets and BDs, we assume a random distribution of inclinations and the 
eccentricity distribution given by \citet{Juric08}. In each simulation, we assigned each target 
a number of planets and BDs from a Poisson distribution, according to the average number 
of planets and BDs per star, calculated from the aforementioned distribution. We also varied 
the outer radius cutoff: 20, 30, 80, 100, 120, 130, and 140 AU. The introduction of an upper 
limit for the planet separation distribution has been suggested by the results of previous 
direct imaging surveys \citep{Chauvin10, Vigan12}.

If a target turns out to have one or more companions in the simulation, we assigned each 
companion a mass and the orbital parameters (semi-major axis, eccentricity, inclination) 
randomly drawn from the assumed distributions. The mass is converted into apparent 
magnitude, given the distance and age of the star and assuming the same family of 
evolutionary models as in Section \ref{sec:data_reduction} \citep[COND][]{Baraffe03}. The 
semi-major axis was converted into a projected separation, given the eccentricity and 
inclination and taking into account the time spent on the orbit. If the combination of 
brightness and separation lies above the contrast curve (\autoref{fig:cc}), then the 
companion is detectable. Thus, at the end of every simulation, we know how many 
companions are created and how many are detected. After 10000 simulations, we 
determined the average detection probability for our A and F main sequence star survey 
(\autoref{fig:prob_dist}). Three companions were detected in this survey, but none of them 
were sub-stellar ($<$80 $M_{\rm Jup}$). Thus, the probability of detecting 0 companions 
for each model is given in \autoref{table:null_result}.

\begin{table}
\caption{Probability of a null result in our A and F main sequence star survey.}
\centering
\begin{tabular}{l c }
\hline 
$r_{cutoff}$ & P(0) [\%] A-type star planet frequency \\
\hline
20 AU & 80 \\
30 AU & 70 \\
80 AU & 25 \\
100 AU & 14 \\
120 AU & 8 \\
130 AU & 6 \\
140 AU & 4 \\
\hline
\end{tabular}
\label{table:null_result}
\end{table}

Given these probabilities, our null result allows us to reject the A-type star model with a 
scaled up planet frequency for $r_{cutoff}>135$ AU, at 95\% confidence. This null result 
is also consistent with previous surveys that found that high mass planets at large orbital 
separations are rare \citep{Nielsen13,Biller13,Desidera15,Chauvin15}. We note 
that unlike larger surveys, our simulations include both the planetary-mass companions 
as well as an extrapolation of the BD companion mass ratio distribution. Since we expect 
a contribution from both populations in the total number of detections, a null result is more 
constraining than a planet population alone. We also adopt a higher planet frequency (as 
expected for A type stars) compared to the standard planet frequency measured by 
\citet{Cumming08} for solar-type stars. If the expected number of detections is higher, a 
null detection result allows one to place more stringent constraints. This is why our survey 
of 13 targets places comparable constraints compared to larger surveys. 

Based on RV measurements of A stars, \citet{Bowler10} suggest positive values for the 
power law indexes of the mass and separation distributions, with high confidence. 
According to our Monte Carlo simulations, if we assume positive power law indexes, the 
probability of a null result is less than 0.1\%, regardless of the planet frequency or the 
radius cutoff assumed. As suggested by \citet{Vigan12}, the inconsistency of direct 
imaging survey results with the distribution parameters from RV observations around 
A-stars \citep{Bowler10} suggests that different planet populations are probed by RV 
measurements at small separations than direct imaging at wide separations.

\section{Conclusions}
\label{sec:conclusion}
We present the results from a survey of carefully selected A- and F-type main sequence 
stars, searching for exoplanet companions. We aim to put direct imaging constraints on 
the occurrence of sub-stellar companions as a function of stellar mass. We obtained data 
on thirteen nearby (d$<$65 pc), young ($<$125 Myr) targets with the APP coronagraph 
on NACO/VLT. We are sensitive to planet masses (2 to 10 M$_{\rm Jup}$) on Solar 
System scales ($\leq$30 AU) for all but one of our targets. We detected a new 
M$6.0\pm0.5$ dwarf companion to HD 984 and confirm stellar companions to HD 12894 
and HD 20385, discovered shortly after our survey data were acquired. Our photometry 
and astrometry for these companions are consistent with the values reported at the time 
of their discovery \citep{Biller13,Hartkopft12}. We found zero false positives in our 
$L'$-band data, as all of our detected point sources were bona-fide companions. We 
perform Monte Carlo simulations to determine the expected probability of detecting 
low-mass companions in our survey, based on our sensitivity and assumed semi-major 
axis distributions. Our non-detection of substellar companions ($<80\,M_{\rm Jup}$) 
allows us to rule out the A-star frequency model distribution for $>$135 AU, with 95\% 
confidence. 

\section*{Acknowledgments}
TM and MAK acknowledge funding under the Marie Curie International Reintegration 
Grant 277116 submitted under the Call FP7-PEOPLE-2010-RG.  Part of this work has 
been carried out within the frame of the National Centre for Competence in Research 
PlanetS supported by the Swiss National Science Foundation. SPQ and MRM 
acknowledge the financial support of the SNSF. EEM acknowledges support from NSF 
award AST-1313029. This paper makes use of the SIMBAD Database, the Vizier 
Online Data Catalog and the NASA Astrophysics Data System.

\bibliographystyle{mn}  

\label{lastpage}

\end{document}